\documentclass{PoS}

\title{The LHCb Vertex Detector Upgrade}

\ShortTitle{LHCb Upgrade}

\author{\speaker{Marina Artuso}\thanks{On behalf of the LHCb VELO group.}\\
         Syracuse University Syracuse NY 13244\\
        E-mail: \email{artuso@physics.syr.edu}}

\abstract{ LHC will offer the opportunity of probing the mass scale of the electro-weak symmetry breaking.
Thus we expect to uncover direct manifestations of physics beyond the Standard Model, which
will raise new questions that may be elucidated by precision
measurements of beauty and charm decays. The LHCb experiment is
poised to pursue this ambitious program as soon as LHC turns on. An
 upgrade to enhance its
physics sensitivity by at least one order of magnitude is critical to the completion of this study, as new physics effects may be subtle. A new
vertex detector is a crucial element of this project.
Important requirements are a radiation resistance up to a fluence of
about 10$^{16}\ n_{eq} {\rm cm}^{-2}$, and a front end electronics capable of  delivering its event information to the back end receiver
boards synchronously with the beam interactions, at 40 MHz. }

\FullConference{17th International Workshop on Vertex detectors\\
		 July 28 - 1 August  2008\\
		 Utö Island, Sweden}

\begin{document}
\newcommand{\Bs}{B_S^0}
\newcommand{\Ds}{D_S^-}
\newcommand{\plus}{+}
\section{Introduction}
Experimental particle physics is at the eve of a very exciting time
that bears to promise of ground breaking discoveries. ATLAS and CMS
are poised to explore the mass scale of the order of the
electro-weak symmetry breaking, and uncover evidence for exotic new
particles that may lead to a better defined path beyond the Standard
Model. LHCb, with its precise and systematic exploration of
interesting beauty and charm decays, will study how the new
particles interfere virtually with $W$ and $Z$ bosons in these
decays. These observations can tell us a great deal about the nature
of new physics, especially their phases.

The present LHCb detector has been designed to be able to cope with
an instantaneous luminosity up to $\sim$ 5$\times 10^{32}$
cm$^{-2}$s$^{-1}$. This is a factor of 20-50 below the design
luminosity of LHC. Thus even without any Super-LHC (SLHC) upgrade,
there is ample room to increase sensitivity, provided that the
experiment can profit from a higher peak luminosity. The first
period of data taking comprises  $\sim$ 10 fb$^{-1}$. The VELO group is
producing replacement modules to be installed during the first data
taking stage in order to make sure that the detector performance will be optimal throughout this whole period. In order
to uncover also very subtle new physics effects, the upgraded LHCb (Super-LHCb) had the goal of
accumulating  $\sim$ 100 fb$^{-1}$ without any detector
replacement \cite{:2008zz} throughout the duration of this second
phase.  Improvement in the hadron trigger algorithm that could increase the sensitivity in
excess of a factor of two are also studied. A fast and efficient reconstruction of the event topology will be a
key element to the success of this approach, and thus
 there is a very close connection between the trigger and the vertex detector R\& D.

\section{The Luminosity Upgrade}
\begin{figure}
\begin{center}
\includegraphics[width=80mm,clip=]{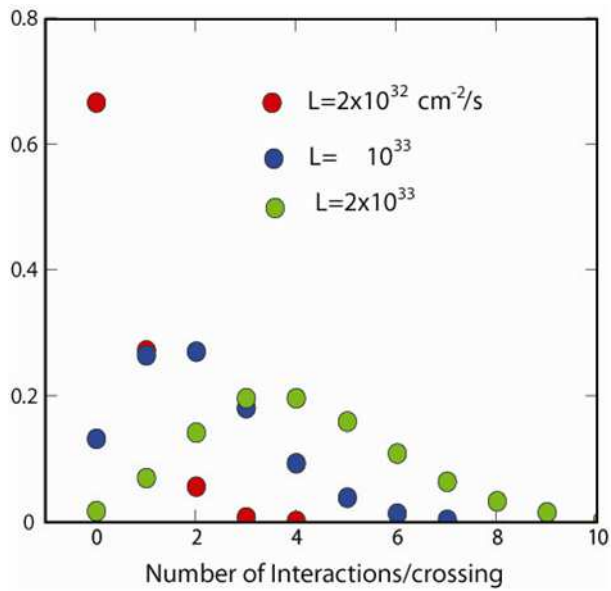}
\caption{Number of interaction per crossing at the nominal LHCb luminosity (2$\times 10^{32}$cm$^{-2}$s$^{-1}$) and at two
different luminosities studied for the LHCb upgrade.}\label{fig:triglumi}
\end{center}
\end{figure}

The bunch crossing rate at the LHCb interaction point (P8) is 40.08
MHz, while 2622 out of the theoretically possible 3564 crossings
have protons in both bunches. Thus the maximum crossing rate with at
least one pp interaction is $\sim$ 30 MHz. Figure~\ref{fig:triglumi}
shows the change in the number of interactions per crossing as a function of the luminosity.
At the present nominal luminosity, about 70\% of the crossings are empty, while the others mostly contain a
single interaction. At a luminosity of 10$^{33}$cm$^{-2}$s${-1}$, there is a dramatic decrease in
the number of empty crossings, but the average number of interactions per crossing is close to 1.
However, already a luminosity of 2$\times 10^{33}$cm$^{-2}$s${-1}$ augments  the average number  of interactions per crossing significantly.
Thus the ability to disentangle topologies associated with multiple events becomes increasingly important to exploit
the higher luminosity.

A very distinctive feature that can be used in designing hadron
triggers that are highly efficient for heavy flavored meson decays
produced in the LHCb event environment while retaining a high
degree of rejection of the more common minimum bias events is the
relatively long decay path of charm and beauty mesons. Detached
vertex criteria are widely used in experiment operating at hadron
machines, but typically they are introduced in the higher trigger
levels. A strategy to incorporate these selection criteria in the
lowest trigger level appears to be a promising avenue to implement
the more robust and general hadron trigger that is a key component
of the LHCb upgrade. This goal requires a strong connection between
the  SUPER-LHCb trigger and   vertex detector design efforts.
The limited granularity of strip detectors is such that an increased occupancy induces a
higher fraction of the so-called ``ghost tracks,'' where a majority of the hits used are not coming from a single well
defined physics track. Ghost tracks are very detrimental to any efficient detached vertex
algorithm. They confuse the vertex topology
and make the vertex reconstruction processors unnecessarily complex and time consuming.

\section{The Present LHCb Vertex Detector System}
\begin{figure}
\begin{center}
\includegraphics[width=160mm,clip=]{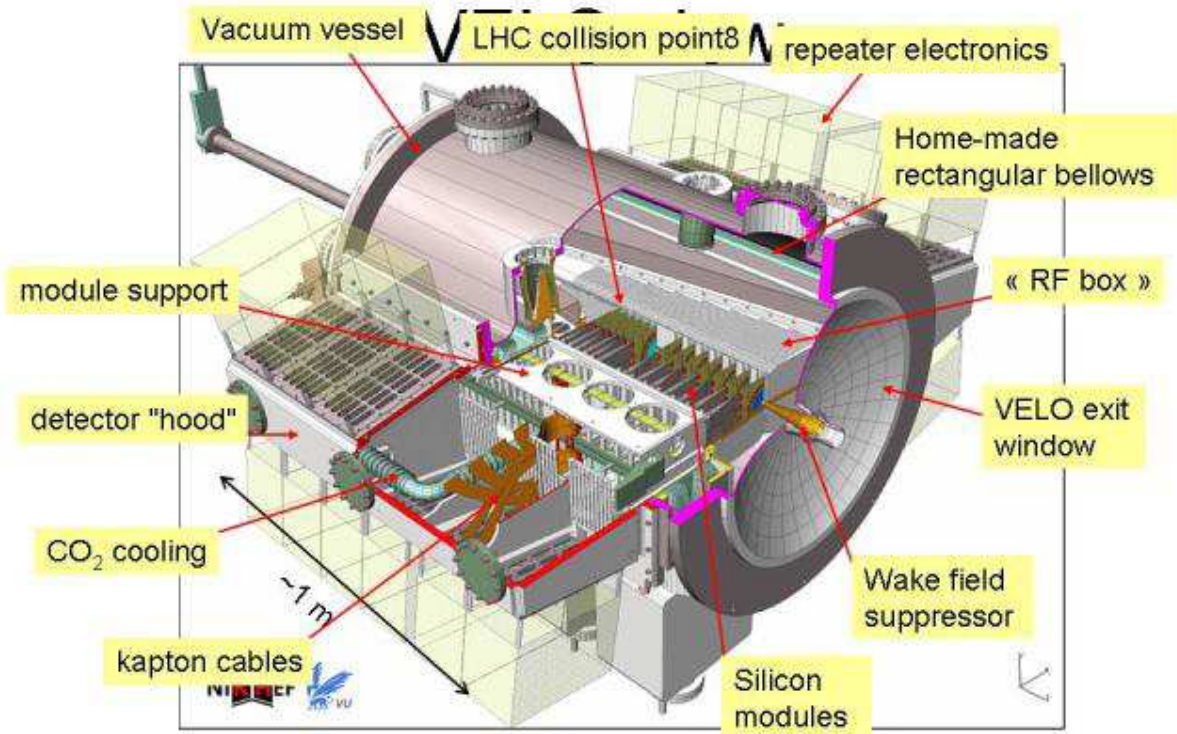}
\caption{Illustration of the VELO detector layout. The detector is
split into tow halves, one of which is shown in this Figure. Each
half comprises 21 modules.} \label{fig:velo}
\end{center}
\end{figure}

The present VErtex LOcator (VELO) silicon detector system consists
of 21 stations of silicon strip detectors positioned along the beam
and perpendicular to the beam axis, as shown in Figure
~\ref{fig:velo}. Each station comprises two double sided modules
that cover 1/2 of the acceptance. They are mounted on retractable
stages, in order to protect the modules from excessive radiation
damage during injection.

Two semi-circular sensors are glued back to back on a hybrid made of
a carbon fiber/TPG (thermal pyrolytic graphite) with kapton flex
circuits laminated on both sides to form a double sided module. The
sensor active area span between 8 and 40 mm. Both of them comprise
2048 strips sensitive either to the radial (R) or azimuthal ($\phi$)
coordinate in a cylindrical coordinate system where Z is along the
beam axis.  More details can be found in Ref.~\cite{:2001hf}. The
hybrids host 16 front end ASICs (BEETLE) \cite{Agari:2004ja} that
include 128 low noise fast analog processors. The analog outputs
are stored in a 4 $\mu$s deep analog pipeline and are read out upon
the arrival of an L0 trigger signal through 4 LVDS lines, each 32
channel deep.

In addition, there are 4 R sensitive modules that are identified as
``VETO stations,'' which have the purpose of rejecting events with
more than 1 interaction. The VETO stations will not be replaced in
the planned upgrade, as we plan to process events with multiple
interactions efficiently.

The LHCb detector configuration, which is oriented in the forward
direction, implies a highly non-uniform irradiation of the silicon
sensors in the vertex telescope.   The innermost radius, being so
close to the beam, will be exposed to $\sim$ 3.4 $\times 10^{14}$ n$_{\rm
eq}/{\rm cm}^{-2}$ per year at the nominal luminosity \cite{gloria-lhcb-note}. Thus the technology chosen for the VELO
sensor is the $n^+$-on-$n$ \cite{non-strips}, adopted by the other
LHC experiments for their inner tracking layers, where $n^+$ strips
are implanted on $n$-type substrate. Interstrip isolation is
achieved with a blanket p-spray technique \cite{p-spray}. Even with
this radiation hard implementation, it is expected that after few
years of operation the VELO detector system will need to be
replaced. Replacement module production is already underway. The
sensor geometry is essentially unchanged, but the sensors are fabricated on
$p$-type substrates. Thus the
detectors are easier to fabricate, as they require only single side
processing and bear the promise of even greater radiation resistance
\cite{ptype:article}.

\section{SUPER-LHCb Vertex Detector Requirements}
The LHCb vertex detector must have fast and robust pattern
recognition capabilities, as it is going to be the heart of the
separation between signal and background events on the basis of the
vertex topology. Further improvements on the proper time resolution
have a significant impact on the detector performance.

One alternative is a reduction of the amount of material in front of
the fist sensing layer in the system. This implies a redesign of the
RF shield, which is also the dominant contributor to the average
radiation length of this detector system. A drastic solution would
be a replacement of the RF shield with wires to screen the beam
image currents. This approach may even enable to  locate the
innermost edge of the modules  closer to the beam axis, with
additional gains in proper time resolution.

Radiation resistance at unprecedented level is a key requirement to maintain  stable operation during the accumulation a $\sim$ 100
fb$^{-1}$ data set. The innermost region
of the detector is expected to receive an integrated fluence of
about 10$^{16}$ 1 MeV n$_{\rm eq}$/cm$^2$, comparable with the
radiation exposure expected for the innermost tracking layers of the
central SLHC detectors.

In order to exploit the higher interaction rate effectively, the front end device must be capable of delivering its information
to back end receiving data boards in real time, namely the hit information should be time stamped and sent to the receiving end synchronously with
the beam crossing frequency of 40 MHz. This requires fast front end electronics, a mechanism to push the data from the sensing element hit to the
digital periphery without excessive delays, and large data rates from the detector to the remote electronics.
This implies impressive data throughput capabilities.

Finally, an overall material minimization for any chosen solution is
necessary to reduce the adverse effects of Multiple Coulomb scattering (MCS). This implies a careful design of the support and
cooling systems, and
an optimization of the number of modules and their geometry.

\section{The VELOPIX Concept}
The approach that has the best chance of satisfying
the requirements described above  relies on hybrid pixel sensors bump bonded
to a custom made front end electronics which satisfies the requirements described above.  This solution allows the a fast
determination of space points that can be aggregated into tracks
with relatively simple pattern recognition algorithms.  In addition,
the use of the pixel geometry would remove the constraint that the
system is perfectly mechanically centered around the beam axis, to
allow the R-z trigger scheme to operate.

As a consequence,
vertex topology algorithms highly efficient for interesting hadronic events can be
implemented in the first step of the selection process. This approach enhances the
sensitivity to $B$ hadronic channels by at least a factor of 2, and is
a selection algorithm that can be successfully exploited at higher luminosities.
Significant  advances towards a first level detached vertex trigger were achieved by the BTeV experiment
\cite{gottshalk}.  The low occupancy, and ease in pattern
recognition featured by vertex detector modules implemented with
hybrid pixel technology are key elements in this approach.

Hybrid pixel devices are our preferred solution because they
represent a mature and robust technology which allows to complete
this project in the relatively stringent time scale envisaged
($\sim$ 2015). This is a mature technology, having been perfected
in the extensive BTeV R\&D and in the development and production of
the ATLAS and CMS vertex detectors. The hybrid approach allows for
independent development and testing of the front end sensors and of
the readout electronics.

There are several advantages associated with the reduced pixel area:
the low input capacitance seen by the preamplifier in the front end
electronics produces a corresponding low intrinsic noise (of the
order of 100 e$^- $), and a negligible common mode noise. This
results in a higher signal to noise ratio than strip devices with
the same detector thickness over a much longer radiation exposure. The extremely low noise occupancy is an
advantage for pattern recognition, as well as the fine two dimensional detector segmentation.

\subsection{VELOPIX Sensor}

In general, sensors with small pixel dimension of approximately 50
$\mu$m are a well developed technology. Two possible improvements to
the standard sensor technology developed for the ATLAS and CMS pixel
devices are currently considered. One of them is the implementation
pixel devices with the $n^\plus$-on-$p$ approach may present
significant advantages in production yields and radiation
resistance.

The closeness of the VELO silicon planes to the LHCb beam makes the
issue of radiation hardness a primary concern. Several different
approaches toward higher radiation sensitive devices are being tried
in the framework of RD50 \cite{rd50}. Prototype of strip and pixel
sensors implemented with a variety of substrates (p-type, n-type
both in float zone and magnetic Czochralski silicon substrates) have
been manufactured, and exposed to different level of irradiation
both from proton and neutron beams. Using the benchmark criteria of
adequate collection properties at the highest levels of radiation
fluence expected, silicon detectors developed on p-type substrates
appear a promising option to achieve these results
\cite{casse:ptype}.

Recent multi-project wafer produced by collaborating institutions
within the RD50 collaboration \cite{rd50} have included pixel
devices with a variety of geometries. In particular, pixel sensors
compatible with the BTeV front end electronics have been designed
with a joint effort between Fermilab and Syracuse University
\cite{artuso:2006}. They have been characterized in the laboratory
before irradiation, with excellent performance. The change in
behavior upon irradiation is currently under investigation.

A solution that alleviates the need for very high (500-1000 V)
reverse bias applied to the silicon devices to achieve adequate
charge collection would provide significantly lower demands on the
high voltage distribution system and on the cooling system. These
``3D detectors,'' constructed with Micro-Electro-Mechanical-Systems
(MEMS) processing, are an innovative approach to segmented solid
state detectors, where narrow holes and trenches are etched through
silicon wafer and the back filled with conductive polysilicon (n and
p doped). The depletion region is achieved with lateral bias, from
electrodes as close as 50 $\mu$m, thus requiring very modest high
voltage. An additional advantage of this solution is the absence of
guard ring,  and the associated dead area. Almost edgeless devices
can thus be implemented. This technology has been proposed  by
Parker and Kenney \cite{parker} in 1994, and it has undergone
several years of intense development. Recent test beam data
\cite{davia} have demonstrated that the technology proposed by
Parker and Kenney can deliver the expected performance. The
challenge is to develop a process for 3D detectors by an industrial
Si device manufacturer. Several efforts are under way \cite{davia},
\cite{fleta}, \cite{dellabetta}.

Finally, the impressive progress made by the RD42 collaboration in
the development of carbon vapor deposition diamond detectors \cite{kagan} suggests that this technology
may be a viable solution for the innermost region of the detector plane, withstanding the highest levels of irradiation.
Poly-crystalline and single crystal chemical vapor deposition (CVD) diamond can
withstand fluences of the order of 10$^16$ n$_{\rm eq}$/cm$^{-2}$, with very competitive signal to noise performance and edgeless
geometry  \cite{kagan}. Their low capacitance
make them naturally a very low noise system, with added noise immunity.  In addition, the leakage current remains
remarkably low at all the levels of irradiation, thus making the cooling system much simpler \cite{rd42}.

\subsection{VELOPIX  Electronics}
The goal to implement zero-suppressed analog readout with correct
beam crossing assignment demands an active program in ASIC
development, to produce a radically different front end device than
the present BEETLE chip. The specifications on its architecture and
data flow are closely connected with the overall data acquisition
and trigger strategy. In particular, the studies of a detached
vertex trigger influence the overall data flow architecture and may
influence some aspects of the front end ASIC design.

\begin{figure}
\begin{center}
\includegraphics[width=80mm,clip=]{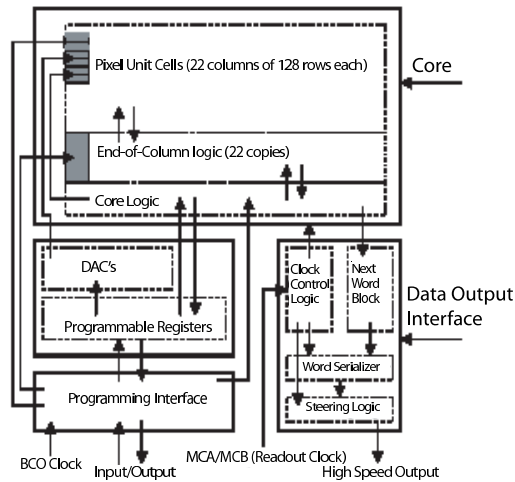}
\caption{Schematic Diagram of the FPIX2 front end ASIC. The analog
processor is an RC-CR preamplifier and shaper followed by a
discriminator to perform zero-suppression. The ``flash'' section
encompasses additional discriminators that implement a 3-bit flash
ADC. The remaining blocks represent end of cell logic and bus
communication with the chip periphery.} \label{fpix2}
\end{center}
\end{figure}

A promising prototype is the FPIX2 chip \cite{fpix2:nim},
developed for the BTeV pixel project. This device consists of an array of 22 columns of 128 cells of 50 by 400 $\mu$m$^2$. Figure~\ref{fpix2} shows a
schematic diagram of the FPIX2 pixel unit cell. It features a RC-CR
preamplifier and shaper, eight comparators which provide zero
suppression and form a 3-bit flash ADC and digital logic. The
digital circuitry encodes the ADC information, and stores the hit
information until an ``output data'' command is received from the
end-of-column logic. Then data are transmitted to the data-output
interface. The built in analog to digital conversion enhances the
robustness of the system, as the information is transmitted in
digital form, and the analog readout provides significant
improvement in the spatial resolution \cite{tb2000}. In particular,
a resolution between 6 and 9 $\mu$m, dependent upon the track angle,
was achieved with a pitch of 50 $\mu$m. The lowest threshold is used
for zero suppression, thus dramatically reducing the amount of
information to be transmitted to the back end electronics. Changes
to adapt this approach to the faster interaction rate and higher
occupancy at LHCb are object of a planned R\&D activity. The device is presently implemented
in radiation tolerant mixed mode TSMC CMOS process. A readout bandwidth of more than 840 MBps has been demonstrated.
It has been irradiated up to 87 MRad, without any degradation of
the analog performance, and with only minor changes on the bias setting \cite{fpix2:nim}.

The group is also studying an adaptation of the TIMEPIX ASIC, developed by the MEDIPIX collaboration, which could allow
the implementation of hybrid pixel systems with a square pixel geometry of roughly $50\times 50\ \mu$m$^2$. This ASIC \cite{timepix}
contains a 256$\times$ 256 matrix of square pixels. Each pixel processors includes a low noise analog front end, with digital readout based on the time over
threshold technique. In the present implementation the pixel data are readout serially through an on-chip LVDS driver. A clock frequency of 100 MHz shifts out the information in
less than 10 ms. A readout architecture consistent with the VELOPIX requirement needs to be implemented. This device is
an evolution of the MEDIPIX2 ASIC, and is implemented in 0.25 $\mu$m radiation tolerant CMOS technology. The new TIMEPIX for VELO is planned as an evolution of the MEDIPIX3
device \cite{medipix3}, and is being designed in $0.13 \mu$m CMOS technology.

\subsection{Hybridization Issues}
The detector electronics hybridization encompasses two main aspects, detector electronics connections, and interconnection between the readout ASIC and the periphery.
A large body of experience on fine pitch bump bonding (down to a minimum pitch of $\sim$ 20 $\mu$m \cite{konig:2007}) has been accumulated in the course of the R\&D and production
of the large pixel devices developed for LHC. Bump bond yields $\sim$ 99.9 \% on known good devices have been achieved. The most outstanding issue worth exploring
with this approach is the minimum thickness that can be implemented both in the sensor and in the readout electronics. As irradiated devices have a collection length of
the order of 100-150 $\mu$m, this thickness is desirable for the sensor elements, while thinning of the readout electronics is limited by operation considerations \cite{lipton}
or mechanical stability issues.

The other connection required, between the hybrid detector ladders and periphery is typically implemented with flex-hybrid circuits, consisting of stacks of metallization layers separated by
think polymid layers. Alternative options are offered by the MCM-D technique \cite{konig:2007}, where thin film technologies are used to deposit dielectric and conductor layers onto the bump side of the sensor layer. One of the advantages of this technique is the elimination of wire bonds, that make the handling of the detector element more challenging. An alternative path is offered
by the ``3D integration,'' which the microelectronic industry consider key to meet the increasing power dissipation, speed, and signal integrity challenges posed by the newest CMOS technologies.
For example, the Fermilab pixel group \cite{lipton} has identified two different commercial processes implementing 3D integration that can be applicable to hybrid pixel devices.
Hybrid modules, including pixel sensors bump bonded to FPIX2 electronics and thinned down to 100 $\mu$m, have been characterized \cite{Ye:2009cc} with electronic test pulses, laser charge injection and radioactive sources. This work shows interesting results, in particular low input capacitance and good interconnection quality. Some of these technologies may allow to implement thinner hybrid modules with better signal integrity.

\subsection{Mechanics and Cooling}
There main issue that is being addressed in the mechanical design of the upgraded detector is the RF shield. On one hand it may need to be adapted to a different detector geometry, on the other hand thinning opportunities are very important as it is crossed prior to any position measurement and thus it has a strong influence on the spatial resolution achievable.
A variety of options are being considered, from different material choices to shapes optimal for the new overall detector geometry.

Finally, cooling plays a critical role in the design of the upgraded detector. In particular, if the innermost portion of the detector is implemented with Si devices, a careful assessment
of whether the present cooling can ensure a temperature that may ensures stable operation without thermal runaway is very important. Studies with different detector models, considering both the
present evaporative CO$_2$ cooling \cite{anne}, and alternative options like the BTeV solution, where pixel modules were mounted on thermal pyrolytic graphite (TPG), which has excellent thermal conductivity, coupled with a liquid nitrogen cooled heat sink \cite{BTeVTDR}.

\section{Conclusions}
The LHCb experiment is at the eve of a very exciting physics program
that will explore several facets of flavor physics with
unprecedented precision. The VELO detector is a key component to
achieve its physics goals.  Radiation resistance and optimal
performance in a hadron trigger in a high occupancy environment are
the performance criteria guiding our R\&D strategy towards the VELOPIX  system.
An essential component of this strategy is the implementation of a front end ASIC that
features low noise readout, zero suppression, and data transmission in real time to the
remote back end electronics that receives the local event fragments and assembles them
in the event record.

\section{Acknowledgements}
This paper benefits tremendously from the work by many colleagues in the VELO Upgrade Group, and from the
extensive discussions with T. Bowcock, P. Collins, C. Parkes, and S.
Stone.
This work was supported by the US National Science Foundation.

\end{document}